\documentclass[12pt,a4paper]{article}
\usepackage{amssymb,amsfonts,amsmath}
\usepackage{color}
\usepackage{epsfig, euscript, graphics}

\begin{document}

\title{Violation of the Bell's type inequalities as a local expression of incompatibility}

\author{Andrei Khrennikov\\ 
Linnaeus University, International Center for Mathematical Modeling\\  in Physics and Cognitive Sciences
 V\"axj\"o, SE-351 95, Sweden}

\maketitle

\abstract{By filtering out the philosophic component we can be said that the EPR-paper was directed against the straightforward interpretation 
of the Heisenberg's uncertainty principle or more generally the Bohr's complementarity principle.
The latter expresses contextuality of quantum measurements: dependence of measurement's output on the complete experimental arrangement. However, Bell restructured the EPR-argument against complementarity to justify nonlocal theories with hidden variables of the Bohmian mechanics' type.  Then this Bell's kind of nonlocality - {\it subquantum nonlocality} - was lifted to the level of quantum theory - up to the terminology  {\it ``quantum nonlocality''}. The aim of this short note is to explain that Bell's  test is simply a special {\it test of local incompatibility of quantum observables}, 
similar to interference experiments, e.g., the two-slit experiment.}

\section{Introduction}

In this note I repeat the message presented in papers \cite{KHBa}-\cite{KHB2}. This note was motivated by the recent INTERNET discussion on the Bell inequality.
 This discussion involved both sides: those who 
support the conventional interpretation of a violation of the Bell type inequalities and those who present a variety of non-conventional positions. As always happens in such discussions on  
the Bell inequality, the diversity of for and against arguments was really amazing, chaotic, and mainly misleading.\footnote{I (as well as R. Gill and 
T. Nieuwenhuizen)  participate in similar discussions since year 2000 and 
the situation did not change so much. It is interesting that the recent (2015) loophole free experiments did not have any impact on ``non-conventional'' part of discussants. But A. Aspect 
told me around year 2000 that closing loopholes would not close the foundational discussion. } 
My attempt to clarify the situation by replying R. Gill generated a new burst of chaos. In parallel, I had the email exchange with W. De Muynck 
and his sharing  my views on the Bell inequality interpretation  stimulated me to write this note (with hope  to stop further misinterpreting  of the Bell inequality).

The aim of this short note is to explain that Bell's  test is simply a special {\it test of local incompatibility of quantum observables}, 
similar to interference  experiments, e.g., the two slit experiment. In particular, Bell's test has nothing to do with nonlocality, neither subquantum nor quantum. I guess that the 
fathers of the Copenhagen interpretation, e.g.,    Bohr and Heisenberg,  would not be interested in such a test  since the principle of complementarity 
was well supported by interference experiments.

The main message of this paper is presented in a few words and formulas in the next section. Only the readers who are deeply involved into quantum foundations can be interested 
in the discussions presented in appendixes.\footnote{ Appendix 1 (coupling  the EPR-argument and the Bohr's complementarity principle, Bohr's reply to Einstein) and in Appendix 2 (the contextual  nature of
the Bohr's principle of complementarity, so to say the principle of {\it contextuality-complementarity}), Appendix 3 (the role of the existence of the irreducible   
quantum of action, the Planck constant).}

\section{CHSH inequality is violated only under the condition of local incompatibility} 
\label{COMP}  

The mathematical counterpart of the story which I plan to tell is well known and rather trivial. The main issue is the interpretation.
Here we proceed by using mathematics from article  \cite{Wolf}. This article is based on the quantum formalism (cf., e.g.,  \cite{KHBa} - \cite{KHB2},  \cite{Muynck}-\cite{BC2}). 
\footnote{The direct appeal to the quantum formalism
makes the presentation clearer for so to say ``real physicists''. Suppose I met a physicist who is not so much interested in quantum foundations, who 
believes in the validity of quantum theory and its completeness, following, for example,  Bohr, Heisenberg, Fock, Landau, and who is not interested in anything 
``beyond quantum mechanics''.  Suppose he asked me:  ``What does a violation of the Bell's inequality mean?''  He is not interested in an answer related 
to hidden variables, or foundations of probability theory.}

\medskip 

Consider the Bohm-Bell  type experiments.  There are considered four observables $A_1, A_2, B_1, B_2$ taking values $\pm 1.$ 
It is assumed that the pairs of  observables $(A_i, B_j), i, j=1,2,$  can be measured jointly, i.e., $A$-observables are compatible with $B$-observables. 
However, the observables in pairs $A_1, A_2$ and $B_1,B_2$ are incompatible, i.e., they cannot be jointly measured. Thus probability 
distributions $p_{A_i B_j}$ are well defined theoretically by quantum mechanics and they can be verified experimentally; 
probability distributions $p_{A_1 A_2}$ and $p_{B_1 B_2}$ are not defined by quantum mechanics and, 
hence, the question of their experimental verification does not arise.  

For spatially separated systems $S_1$ and $S_2,$ incompatibility of the $A$-observables on $S_1$
and the $B$-observables on $S_2$ is natural to call {\it local incompatibility}

Now we proceed generally, i.e., without the local incompatibility assumption. 
In the quantum formalism observables are presented by Hermitian operators which will be denoted  by the same symbols as the observables.
The quantum representation of the CHSH-correlations can be obtained with the aid of the ``CHSH-operator'':
\begin{equation}
\label{L1}
C= \frac{1}{2}[A_1(B_1+B_2) +A_2(B_1-B_2)].
\end{equation}
As was shown in article \cite{Wolf}, for given observables,  the CHSH inequality holds for all quantum states iff
\begin{equation}
\label{L1}
C^2 \leq I,
\end{equation}
where $I$ is the unit operator.

It is easy to show that 
\begin{equation}
\label{L2}
C^2=1+(1/4) [A_1,A_2][B_1,B_2].
\end{equation}

Thus if {\it at least one of the commutators} equals to zero, i.e., 
\begin{equation}
\label{L3}
[A_1,A_2]=0, 
\end{equation}
or
\begin{equation}
\label{L4}
 [B_1,B_2]=0,
\end{equation}
then the CHSH inequality holds for all quantum states. 
It is impossible to violate this inequality, if at least at one side (say at the $A$-side) observables are compatible. 
We emphasize that {\it both these conditions are local.} Condition (\ref{L3}) is about observables on  system $S_1$  and 
condition (\ref{L3}) is about observables on system $S_2.$  (How can one find any trace of nonlocality here?) 

\section{Conclusion}

In the light of the previous consideration, it is clear that the common claim that the Bell story about nonlocality is really misleading. 
(Any rationally thinking expert in quantum foundations should  dismiss the use of the term ``nonlocality''.)
The Bell-tests are just special tests of incompatibility.  The natural question arises: 

\medskip

{\it Do the Bell type experiments add some foundational value to the interference type experiments?}   
 
 \medskip
 
Thus if one hope to go beyond quantum mechanics, then he  should  {\it  search for possible violations of the complementarity principle. }
However, such experiment have nothing to do with spatial separation of systems.  
  
 \section*{Appendix 1: The role of complementarity principle in the EPR-reasoning}
 
I start with the remark that one has to separate the Bell inequality as the mathematical result and its interpretation, including conditions of its derivation. So, there is no problem 
with the ``Bell theorem'', the problem is the conventional interpretation of this mathematical statement.  Thus people who claim that they found mistakes in  Bell's mathematics do very 
bad job for quantum foundations: they discredit the rational critique of the conventional interpretation of the Bell inequality. 

The second important remark is that using the notion {\it ``local realism''} in interpreting violations of the Bell type inequalities  is really misleading. 
It seems that Bell have never used this notion. At least in his book \cite{BE2}, the collection of his main works, ``local realism'' appears only in the title 
of the reference to the of one philosopher. 

Bell started his project as being on the realist position \cite{BE2} (by following directly the EPR-paper \cite{EPR}). Then he claimed that realism can be saved only through the rejection 
of locality. Finally, he understood that 

\medskip

{\it ``the result of an observation may reasonably depend not only upon the state of the system (including the hidden variables)
but also on the complete disposition of the apparatus''}  \cite{B14}. 

\medskip

Shimony \cite{SHIM} stressed that this is the first statement about contextuality (although Bell did not use this terminology):

\medskip

{\footnotesize  ``John Stewart Bell (1928-90) gave a new lease on life to the program of hidden variables by proposing contextuality. In the physical example just considered the
complete state $\lambda$  in a contextual hidden variables model would indeed ascribe an antecedent element of physical reality to each squared spin component $s_n^2$ but in a
complex manner: the outcome of the measurement of $s_n^2$ is a function $s_n^2(\lambda, C)$ of the hidden variable $\lambda$ 
and the context $C,$ which is the set of quantities measured along with $s_n^2.$ ... a minimum constraint on the context $C$ is that it consist
of quantities that are quantum mechanically compatible, that is represented by self-adjoint operators which commute with each other. ...''}

\medskip

However, this is nothing else as Bohr's statement that measurement's output depends on the complete 
experimental arrangement (see Appendix 2). In principle, there is just one step from understanding of the role of such type of contextuality t
o Bohr's complementarity principle and to understanding 
that violations of the Bell-type inequalities is a consequence of this principle (Appendix 2). Unfortunately, Bell was addicted to the nonlocality issue and this final step has never been done.  

One may say that Bohr did not consider the possibility to go beyond the quantum theory, e.g., by introducing hidden variables. As I learned from Arkady Plotnitsky who is  the author of a series 
of monographs about Bohr, the complementarity principle, and the interpretations in the {\it Spirit of Copenhagen} (see, e.g.,  \cite{PL1, PL2} and our joint paper \cite{PLKHR}), 
in principle Bohr did not reject the possibility of construction of mathematical subquantum models. He was not excited by mathematical no-go theorems.
For him the possibility to go beyond quantum theory was blocked by the principle of complementarity. Some people think that the latter is a purely philosophical principle.
However, for Bohr and other Copenhagenists, this principle was a physical principle such as Newton laws (see Appendix 2).   

Bell's behavior has its root in the EPR-paper \cite{EPR} in that the anti-complementarity issue was covered by  the ``philosophic souse'' with ``elements of reality'' as its basic 
component. Let us clean the EPR-argument from the ``philosophic souse''. Such a rafined argument has the following structure: 
\begin{itemize}
\item The Copenhagen interpretation is heavily based on the complementarity principle.\footnote{It seems that in the EPR-paper it was identified 
with Heisenberg's uncertainty principle. But at that time  such identification was common. }
\item This principle was (straightforwardly) interpreted as the rejection of the joint {\it existence} of the position and momentum of a quantum system.
\item The existence of the EPR-states  exhibiting perfect correlations contradicts
to the   complementarity principle.
\end{itemize}
The natural conclusion from the EPR-argument could be formulated as follows:

\medskip

{\it Bohr's principle of complementarity should be rejected!}

\medskip

However, Einstein, Podolsky, and Rosen combined the above reasoning on physical foundations of quantum mechanics with philosophic argumentation based on 
the elements of reality. The above conclusion was not clearly stated in \cite{EPR}. 

This is the good place to mention the Bohr reply \cite{BR} to the EPR-paper.  During conferences on quantum foundations, speakers permanently say that it is impossible 
to understand  Bohr's paper. I do not think so. The message of this paper is very clear:

\medskip
 
{\it The principle of complementarity is not about the joint existence. It is about the possibility of joint measurement!} 
  
\medskip  

In any event, Bell did not understand the Bohr reply and tried to justify the EPR-argument by using subquantum mathematical models of the  hidden  
variables type. As was already stressed, ignoring the crucial role played by the complementarity principle in the EPR-paper led him to the emphasis 
of nonlocality. We recall that in the EPR-paper \cite{EPR} nonlocality was mentioned as an absurd alternative to incompleteness  of quantum mechanics.
(And we interpret the latter as the possibility to violate  the complementarity principle.)

\section*{Appendix 2: Bohr's complementarity principle as a principle of contextuality}  

Bohr's complementarity principle is not just a kind of the impossibility principle. 
In the modern terminology the essence of the  complementarity principle is contextuality. But Bohr by himself did not operate with the notion ``experimental context''; he wrote 
about experimental conditions \cite{BR0}:

\medskip  

 {\footnotesize ``Strictly speaking, the mathematical formalism of quantum mechanics and electrodynamics
merely offers rules of calculation for the deduction of expectations pertaining
to observations obtained under well-defined experimental conditions specified
by classical physical concepts.''} 
 
\medskip

We now present the basic futures of Born's principle of contextuality-complementarity \cite{KHROP}:
  
\begin{itemize}
\item (B1): There exists the fundamental quantum of action given by the Planck constant $h.$
\item (B2):  The presence of $h$ prevents approaching internal features of quantum systems.
\item (B3): Therefore it is meaningless (from the viewpoint of physics) to build scientific theories about such features.
\item  (B4): An output of  any observable   is composed of contributions from a system under measurement and the measurement device.
\item (B5): Therefore the complete experimental arrangement (context) has to be taken into account.
\item (B6): There is no reason to expect that all experimental contexts can be combined. Therefore there is no reason to expect that all observables can be measured jointly.
This is the essence of {\it the principle of complementarity.} Hence, there exist incompatible observables.   
\end{itemize}

\medskip

We emphasize that the principle (B6) is just a  consequence of (B4) and (B5). And Bohr understood the complementarity principle as the combination of (B1-B6). 
It is the good place to remark that statement (B6) is very natural in any experimental situation. The existence of incompatible experimental contexts is not surprising.
Non-existence would be really surprising.
 
Unfortunately, the name ``complementarity principle'' led to the common understanding of this principle as just an impossibility prtinciple. 
I think that in the light of the above consideration, the name of Bohr's ``complementarity principle'' should be changed to 

\medskip

{\it The Bohr's contextuality principle or the contextuality-complementarity principle.} 

\medskip

Of course, such renaming would be a delicate operation since nowadays the terminology ``contextuality'' is widely used  in foundational discussion on the Bell type inequalities.
In such discussions, the meaning of the notion ``contextuality'' does not coincide with so to say ``Bohr's contextuality'', as taking into account the experimental context  
to explain the mechanism of generating the values of quantum observables.  From this viewpoint, any single quantum observable is contextual. Roughly speaking contextuality 
is present everywhere. There is no need  in consideration joint measurements of compatible observables, $A, B_1$ and $A, B_2,$ to find its exhibitions.  

\section*{Appendix 3: A role of the Planck quantum of action}

As was already noted, the Bohr's reasoning leading to the complementarity principle (B6) was fundamentally based on the principle (B1): 
the existence in nature of the fundamental quantum of action given by the Planck constant $h.$ It seems that nowadays  this fundamental ``Bohr-Planck postulate'' is totally forgotten. 
 I have not seen any its trace in nowadays foundational discussions on the complementarity principle. Moreover, the Planck constant is not 
present at all in modern discussions: neither  the Bell's type no-go theorems nor theorems on  ``quantum contextuality'' do not containn $h$
(nor the Kochen-Specker theorem). I think that the essence of the complementarity principle (and hence the role of the EPR-counterargument) 
can be understood only through the careful analysis of the Bohr-Planck postulate. Such an analysis with attraction of new ideas can be done only by 
real physicist. The mathematical games around no-go theorem would not help so much. 
In principle, we cannot exclude that Bohr was wrong and the existence of the fundamental quantum of action does not imply contextuality-complementarity. 
However, if he were right, then, as we have seen in section \ref{COMP}, all manipulations with classical probabilistic constraints have not so much relation 
to physics.

 \end{document}